\documentclass[aps,prd,twocolumn,groupedaddress,showpacs,showkeys,nofootinbib]{revtex4}%
\usepackage{mathrsfs} 
\usepackage{graphicx}
\usepackage{bm}
\usepackage{amssymb,amsmath,amsthm}
\begin{document}%
\title{\bf Lorentz symmetry breaking as a quantum field theory regulator}
\author{Matt Visser}
\affiliation{School of Mathematics, Statistics, and Operations Research,
Victoria University of Wellington, Wellington 6140, New Zealand}
\date{26 April 2009; 30 June 2009;
\LaTeX-ed \today}
\begin{abstract}
Perturbative expansions of quantum field theories typically lead to ultraviolet (short-distance) divergences requiring regularization and renormalization. Many different regularization techniques have been developed over the years, but most regularizations require severe mutilation of the logical foundations of the theory. In contrast, breaking Lorentz invariance, while it is certainly a radical step, at least does not damage the logical foundations of the theory. I shall explore the features of a Lorentz symmetry breaking regulator in a simple polynomial scalar field theory, and discuss its implications. In particular, I shall quantify just ``how much'' Lorentz symmetry breaking is required to fully regulate the quantum theory and render it finite. This scalar field theory provides a simple way of understanding many of the key features of Petr Ho\v{r}ava's recent article [Phys.\ Rev.\ {\bf D79} (2009) 084008] on 3+1 dimensional quantum gravity.
\end{abstract}
\keywords{Lorentz symmetry; regularization; renormalization; finite QFTs}
\pacs{11.30.Cp 03.70.+k   11.10.Kk   11.25.Db    04.60.-m}
\maketitle
\def\lint{\hbox{\Large $\displaystyle\int$}} 
\def\hint{\hbox{\Huge $\displaystyle\int$}}  

\def\d{{\mathrm{d}}}
\newcommand{\scri}{\mathscr{I}}
\newcommand{\sun}{\ensuremath{\odot}}

\section{Introduction}

The ultraviolet divergences that typically infest perturbative expansions of relativistic quantum field theories have been a topic of interest (and sometimes heated debate) for over 60 years~\cite{Dirac}. As a practical matter, loop integrals in Feynman diagrams often lead to ultraviolet divergences requiring at the very least some form of regularization, typically followed by renormalization~\cite{Schweber, Nishijima, Ramond, Rivers, Weinberg}. Many different regularization techniques have been developed over the years. Most regularizations are designed primarily for computational efficiency, and retain as many symmetries as possible to simplify the algebra --- even if this requires (at least in intermediate stages of the computation)  severe mutilation of the logical foundations of the theory. 

For example: Pauli--Villars regularizations violate unitarity, so that probabilities are not necessarily positive at intermediate steps of the calculation.  Similarly Lorentz-invariant higher-derivative regularizations also violate unitarity; while dimensional regularization involves a fictitious analytic extension to non-integer dimensions, (and when fermions are added, this must be coupled with an equally fictitious analytic extension of the Dirac gamma matrix algebra to non-integer dimensions). It seems that retaining Lorentz invariance almost guarantees that the regulator must break something even more fundamental in the theory. (With the only possible exception being the ``finite field theories'' based on supersymmetry --- such as $N=4$ SUSY Yang--Mills~\cite{Stanley}, and $N=8$ supergravity~\cite{Zvi, Kallosh}, and their variants --- though even there one typically resorts to unphysical regulators at intermediate stages of the calculation.)

In contrast, breaking Lorentz invariance, while it is certainly a radical step, does not damage the logical foundations of the theory.  It is an \emph{experimental observation} that empirical reality obeys Lorentz symmetry to very high accuracy, but it is not a \emph{logical necessity}.  Breaking Lorentz invariance to keep the quantum field theory finite may lead to complicated algebra, but at least it does not undermine the logical and physical foundations of the theory.  It then becomes an \emph{empirical} question as to whether or not the theory is ultimately compatible (either for finite cutoff, or in the limit as the cutoff is sent to infinity) with the observed bounds on Lorentz symmetry violations~\cite{Mattingly,  Kostelecky-lsb,  Kostelecky-conferences, Liberati,Ellis, Bailey, Albert, Katori}.

 In this article I shall explore the features of a Lorentz symmetry breaking regulator in a simple polynomial scalar field theory, and discuss its implications. In particular, I shall precisely quantify just ``how much'' Lorentz symmetry breaking is required to fully regulate the scalar field theory and render it finite. As an application, I shall then show how this model provides a simple way of understanding many of the key features of Ho\v{r}ava's recent article~\cite{Horava} on 3+1 dimensional quantum gravity.

\section{Free Lagrangian}
\label{S:free}

In flat $d+1$ spacetime consider the action:
\begin{equation}
S_\mathrm{free} = \int   \left\{ \dot \phi^2 -  \phi (-\Delta)^z \phi \right\} \d t \; \d^d x. 
\end{equation}
Here $\Delta = \vec\nabla^2$ is the spatial Laplacian, and $z$ is some positive integer.  (Lorentz invariance corresponds to $z=1$. This model explicitly preserves both parity and ordinary spatial rotational invariance.) We have used the theorists' prerogative to choose units such that the coefficient of the time derivative term equals the coefficient of the spatial derivative term --- this is in contrast to the usual choice of setting $c=1$. We shall certainly have $c\neq1$ in the present proposal, and have instead set the coefficent of $\Delta^z$ to unity to simplify the power counting. We also set $\hbar\to 1$.  

(If one is worried about adopting this particular choice of ``theoretician's units'',  one can always go to the more standard ``physical units'' ($c=1$); see section~\ref{S:physical} for details. Doing so will only serve to make the details somewhat messier but will lead to no new physics.) 

With this choice of units, consider the engineering dimensions (canonical dimensions) of space and time: We immediately deduce
\begin{equation}
[\partial_t] = [\vec\nabla]^z; \qquad 
[\d t] = [\d x]^z.
\end{equation}
But since we want the action to be dimensionless
\begin{equation}
[S]=[1],
\end{equation}
we see that
\begin{equation}
 [\phi] = [\d x]^{(z-d)/2}.
\end{equation}
This immediately suggests that the case $z=d$ will play a very special role in the discussion, since the field $\phi$ is then dimensionless.

It is convenient to define formal symbols $\kappa$ and $m$ having dimensions of momentum and energy
\begin{equation}
[\kappa] = 1/[\d x],   \qquad   [m] = 1/[\d t],
\end{equation}
since then
\begin{equation}
[m] = [\kappa]^z,
\end{equation}
and
\begin{equation}
 [\phi]  = [\kappa]^{(d-z)/2} = [m]^{(d-z)/(2z)}.
\end{equation}
Note that for Lorentz invariance, $z=1$, we recover the usual result $[\phi]=[m]^{(d-1)/2}$, so that in particular $\phi$ is dimensionless in 1+1 dimensions, $\phi$ has dimensions of (mass)${}^{1/2}$ in 2+1 dimensions,  $\phi$ has dimensions of mass in 3+1 dimensions, and $\phi$ has dimensions of (mass)$^2$ in 5+1 dimensions. These are the usual and expected results.

Now add the various possible sub-leading terms to this free Lagrangian
\begin{equation}
S_\mathrm{free} = \int   \left\{ \dot \phi^2 -  \phi \left[\,m^2 - c^2 \Delta + \dots + (-\Delta)^z\, \right] \phi \right\} \d t \; \d^d x.
\end{equation}
Note that now
\begin{equation}
[c] = [\d x/\d t] = [\d x]^{1-z} = [\kappa]^{z-1}=  [m]^{(z-1)/z},
\end{equation}
which is why, (given the other choices we have already made above), we do \emph{not} have the freedom to set $c\to 1$, (unless of course $z=1$). Note that these sub-leading terms all have positive momentum dimension (and positive energy dimension) --- treated perturbatively, we shall soon see that they correspond to super-renormalizable operators.

\section{Interactions}
\label{S:power-count}
Now add polynomial interactions:
\begin{equation}
S_\mathrm{interaction} = \int   P(\phi)  \; \d t \; \d^d x 
 = \lint \left\{ \sum_{n=1}^N  g_n \, \phi^n \right\} \d t \; \d^d x.
\end{equation}
We shall refer to the resulting quantum field theory, defined by $S=S_\mathrm{free}+S_\mathrm{interaction}$, as $P(\phi)^z_{d+1}$. 
Each coupling constant $g_n$ has engineering dimensions
\begin{equation}
[g_n] = [\kappa]^{d+z- n(d-z)/2} =  [m]^{[d+z- n(d-z)/2]/z}.
\end{equation}
So the couplings have non-negative momentum dimension (and so also have non-negative energy dimension) as long as
\begin{equation}
d+z- {n(d-z)\over 2} \geq 0.
\end{equation}
Recalling that $d$, $z$, and $n$ are by definition all positive integers, this is equivalent to either
\begin{equation}
n \leq {2(d+z)\over d-z}; \qquad (\hbox{provided } z < d),
\end{equation}
or
\begin{equation}
n \leq \infty; \qquad (\hbox{provided } z\geq d).
\end{equation}
This is enough to imply that the theory has the correct  ``power counting''  properties to be renormalizable. 
Indeed, based on our intuition from studying Lorentz invariant theories~\cite{Ramond, Rivers, Weinberg} this very strongly suggests, (and appealing to the technical results of Anselmi and Halat~\cite{Anselmi} we shall quickly verify), that the theory is renormalizable as long as the highest power $N$ occurring in the polynomial $P(\phi)$ is either
\begin{equation}
N = {2(d+z)\over d-z};  \qquad (\hbox{provided } z < d),
\end{equation}
or 
\begin{equation}
N = \infty; \qquad (\hbox{provided } z\geq d).
\end{equation}
For Lorentz invariance, $z=1$, this reduces to 
\begin{equation}
N_{(z=1)} ={2(d+1)\over(d-1)}.
\end{equation}
This is completely compatible with the usual standard results: $\phi^n$ is renormalizable for any positive integer $n$ in 1+1 dimensions; $\phi^6$ is renormalizable in 2+1 dimensions, $\phi^4$ is renormalizable in 3+1 dimensions, and $\phi^3$ is renormalizable in 5+1 dimensions.
Note that there is something (reasonably elementary) to verify regarding this dimensional analysis argument --- to check convergence of the Feynman diagrams we need a minor generalization of the usual argument characterizing the ``superficial degree of divergence". It is at this stage that we need to appeal to the technical results of Anselmi and Halat~\cite{Anselmi}. 

\section{Superficial degree of divergence}
\label{S:superficial}

Consider a generic Feynman diagram. As usual, for each loop in the truncated Feynman diagram we pick up an integral~\cite{Ramond, Rivers, Weinberg}
\begin{equation}
\int \d \omega_\ell \; \d^d k_\ell \dots
\end{equation}
In contrast, for each internal line we now pick up a propagator $G(\omega,\vec k)$~\cite{Ramond, Rivers, Weinberg}  that violates Lorentz invariance:
\begin{equation}
{1\over (\omega_\ell-\omega_e)^2 - \left\{m^2 + c^2(\vec k_\ell - \vec k_e)^2 +\dots + [(\vec k_\ell-\vec k_e)^2]^z \right\} }.
\end{equation}
Here $\omega_e$ and $\vec k_e$ are some linear combination of the external momenta, and $\omega_\ell$ and $\vec k_\ell$ are the loop energy and loop momentum respectively. Let $L$ be the number of loops, and $I$ the number of internal propagators.  Then each loop integral has dimension
\begin{equation}
\int \d \omega \; \d^d k \to [\d\omega] [\d k]^d = [\kappa]^{d+z},
\end{equation}
while each propagator has dimension
\begin{equation}
G(\omega,\vec k)  \to [\kappa]^{-2z}.
\end{equation}
Thus for the entire Feynman diagram the total contribution to dimensionality coming from loop integrals and propagators is
\begin{equation}
 [\kappa]^{(d+z)L-2Iz},
\end{equation}
which is summarized by saying that in this Lorentz-violating situation the ``superficial degree of divergence'' is
\begin{equation}
\delta = (d+z)L-2Iz.
\end{equation}
If $z=1$, the Lorentz invariant situation, this reproduces the standard result $ \delta = (d+1)L-2I$. See, for example, Ramond~\cite{Ramond}, page 139, equation (2.2), or Rivers~\cite{Rivers}, page 45,  equation (3.8). See also the articles by Anselmi and Halat~\cite{Anselmi} for more details on the superficial degree of divergence for Lorentz-violating theories.

We can rewrite the general (Lorentz-violating) result as
\begin{equation}
\delta = (d-z)L-2(I-L)z.
\end{equation}
But to get $L$ loops one needs, at the very least, $I$ propagators. So for any Feynman diagram we certainly have
\begin{equation}
\delta \leq  (d-z)L.
\end{equation}
It is a standard result that if the superficial degree of divergence is negative, and the superficial degree of divergence of every sub-graph is negative, 
then the Feynman diagram is convergent~\cite{Ramond, Rivers, Weinberg}. (See also Anselmi and Halat~\cite{Anselmi}.)
Consequently, if one picks $d=z$ then for any Feynman diagram
\begin{equation}
\delta \leq  0,
\end{equation}
and the \emph{worst} divergence one can possibly encounter is logarithmic. (Or a power of a logarithm if one has several subgraphs with $\delta=0$.) Such a logarithmic divergence can occur only for $L=I$, that is for a ``rosette'' Feynman diagram.  This observation is enough to guarantee that  $P(\phi)^{z=d}_{d+1}$ is power-counting renormalizable. 

In fact considerably more can be said:  
Since ``rosette'' Feynam diagrams can simply be eliminated by normal ordering, it follows that  the normal-ordered theory denoted  by $:\!P(\phi)^{z=d}_{d+1}\!:$ \, is power-counting ultraviolet finite. 

Indeed, for $d=1$, and hence $z=1$, eliminating the ``rosette'' Feynman diagrams via normal ordering is, as per Simon's book~\cite{Simon}, the key technical ingredient to proving the perturbative finiteness of the normal-ordered $:\!P(\phi)^{z=1}_{1+1}\!:$ field theory. See also Glimm and Jaffe~\cite{Glimm-Jaffe} for a similar discussion.

Furthermore, if $z>d$ then there are no superficially divergent Feynman diagrams whatsoever, and the entire theory is power-counting finite.
That is, for $z>d$ one does not even need to bother normal ordering the $P(\phi)^{z>d}_{d+1}$ field theory in order to get something  power-counting ultraviolet finite. Now combining these power-counting arguments with the technical machinery developed by Anselmi and Halat~\cite{Anselmi}, I emphasise that the two central technical results of the present article are:
\begin{itemize}
\item 
With normal ordering,  $:\!P(\phi)^{z=d}_{d+1}\!:$ is perturbatively ultraviolet finite.
\item
Even without normal ordering, $P(\phi)^{z>d}_{d+1}$ is perturbatively ultraviolet finite.
\end{itemize}

\section{Physical ($c=1$) units}
\label{S:physical}

Suppose we instead adopt the more usual ``physical'' units where $c\to 1$, in that case we would write the propagator $G(\omega,\vec k)$ as
\begin{equation}
{1\over \omega^2 - \left\{m^2 + (\vec k\,)^2 +\dots + \zeta^{2-2z}[(\vec k\,)^2]^z \right\} },
\end{equation}
where $\zeta$ is now a parameter with the physical units of momentum that controls the scale of Lorentz symmetry breaking, and in this section we now have
\[
[\zeta] = [\kappa] = [m].
\]
Now introduce an explicit momentum cutoff $\Lambda$ for the loop integral. In these ``physical'' units the appropriate energy cutoff is then $\Omega = \zeta^{1-z} \Lambda^z$, and for each loop
\begin{equation}
\int \d \omega \; \d^d k \to \Omega \; \Lambda^d = \zeta^{1-z} \; \Lambda^{d+z},
\end{equation}
This asymmetric cutoff in the loop integration is absolutely essential; a condensed matter physicist would say that we are considering a system subject to ``anisotropic scaling''.
Furthermore,  for each propagator
\begin{equation}
G(\omega,\vec k)
\to \zeta^{2z-2} \; \Lambda^{-2z}.
\end{equation}
Thus for the entire Feynman diagram the cutoff dependence is
\begin{equation}
 \zeta^{(z-1)(2I-L)}\;\; \Lambda^{(d+z)L-2Iz},
\end{equation}
which is again summarized by saying that the ``superficial degree of divergence'' is
\begin{equation}
\delta = (d+z)L-2Iz.
\end{equation}
This is the same result as perviously obtained using the ``theoretician's units'' of sections~\ref{S:free}--\ref{S:power-count}--\ref{S:superficial}, as of course it must be. Some readers may prefer this point of view, (especially when trying to compare results between the particle physics and condensed matter sub-disciplines), but the ultimate physics results cannot be affected by this change of units.

\section{3+1 dimensions}
\label{S:3+1}

In the specific case of 3+1 dimensions it is sufficient to consider $z=3$, and so up to six spatial derivatives. That is, in ``theoreticians' units'', take $S_\mathrm{free}$ to be
\begin{equation}
 \int   :\!\left\{ \dot \phi^2 -  \phi \left[\,m^2 - c^2 \Delta + \Xi^2 \;\Delta^2 + (-\Delta)^3 \,\right] \phi \right\}\!: \d t \; \d^3 x,
\end{equation}
and take $S_\mathrm{interaction}$ to be
\begin{equation}
 \int   :\!P(\phi)\!:  \; \d t \; \d^3 x 
 = \lint \left\{ \sum_{n=1}^\infty  g_n \; :\!\phi^n\!: \right\} \d t \; \d^3 x.
\end{equation}
Then
\begin{equation}
[\phi] = [1];   \qquad [g_n]= [m^2] = [\kappa^6],
\end{equation}
and so the scalar propagators $G(\omega,\vec k)$ are sixth-order polynomials in spatial momentum, of the form
\begin{equation}
{1\over \omega^2 - \left\{m^2 + c^2(\vec k)^2 +\Xi^2 \; [(\vec k)^2]^2  + [(\vec k)^2]^3 \right\} },
\end{equation}
or more formally
\begin{equation}
{1\over \omega^2 - \left\{m^2 + c^2\; k^2 +\Xi^2 \; k^4  + k^6\right\} },
\end{equation}
The key point here is that the field $\phi$ is dimensionless.
By our general argument, this quantum field theory is by construction perturbatively ultraviolet finite.

\section{Why now?}

Why has this not been done before? There is a mixture of reasons: A key point is that when breaking Lorentz invariance explicit loop calculations become computationally difficult. This feature has to be balanced against the fact that one is adopting a regulator that is ``physical'' --- in the sense that the regulated theory makes perfectly good sense as a quantum field theory in its own right. One does not have to perform any delicate limiting procedure to recover a logically consistent quantum field theory.

Historically, Lorentz symmetry violations have typically been viewed as either non-existent, or as renormalizable perturbative additions to an otherwise Lorentz symmetric theory~\cite{Kostelecky}.  For early work suggesting a breakdown of Lorentz symmetry at high energies, see~\cite{Pavlopoulos}.  For some recent work on quantum field theories exhibiting Lorentz symmetry breaking, see~\cite{Anselmi}. Note that  Anselmi and Halat's notion of  ``weighted power counting''~\cite{Anselmi} is essentially identical to the Lorentz-violating extension of the usual notion of ``superficial degree of divergence'' discussed above.

In those situations one has to worry about the question of whether or not Lorentz violating terms that naively seem to dominate only at high energies might somehow, through loop diagrams, contaminate the low-energy physics and lead to significant fine tuning problems~\cite{Sudarsky}.  There are contrasting opinions to the effect that in many situations low-energy Lorentz symmetry is a fixed point of the renormalization group, which might to some extent ameliorate detectable manifestations of Lorentz symmetry breaking~\cite{Holger}. 

These issues are less of a concern in the current approach: Since the regularized normal-ordered Lorentz-violating quantum field theory is actually finite, (the few remaining logarithmic divergences being cured by the normal ordering), we can safely use the tree-level action as a reasonable approximation to the full effective action.  In the low-momentium limit, the lowest-momentum terms will dominate and the propagators are effectively of the form
\begin{equation}
G(\omega,\vec k)  \to 
{1\over (\omega_\ell-\omega_e)^2 - \{m^2 + c^2(\vec k_\ell - \vec k_e)^2 \} },
\end{equation}
which is a Lorentz invariant dispersion relation, thereby indicating the low-momentum recovery of Lorentz invariance as an accidental symmetry.  At one level this can be related to the observation by Holger Nielsen \emph{et al.}~\cite{Holger}, that Lorentz symmetry breaking terms are often suppressed in the low-momentum limit, but there is a more instructive observation that one can draw from condensed matter and atomic/ molecular/ optical [AMO] physics: There are many physical systems in which the perturbations/quasi-particles are described by the Bogoliubov dispersion relation~\cite{LSB}, which in its most general form is described by
\begin{equation}
\omega(\vec k) = \sqrt{ m^2 + c_s^2\; (\vec k)^2 +\Xi^2\; [(\vec k)^2]^2  },
\end{equation}
or more schematically by
\begin{equation}
\omega(k) = \sqrt{ m^2 + c_s^2\; k^2 +\Xi^2\; k^4  }.
\end{equation}
If $m=0$ then at low momenta the $c_s^2\; k^2$ term dominates, and one obtains phonons travelling at the speed of sound $c_s$. At high momenta, the $\Xi^2\; k^4 $ term dominates and one recovers a non-relativistic spectrum for the quasiparticles. In the language of ``anisotropic scaling'' working with the Bogoliubov dispersion relation corresponds to working at a $z=2$ ``Lifshitz point''. Indeed, in various explicit computations related to the ``analogue spacetime'' programme~\cite{analogue}, computations in which we were concerned with the response of otherwise-free quasi-particle QFTs when subjected to external constraints~\cite{ameliorate}, we have encountered situations where the $\Xi^2\; k^4$ term in the Bogoliubov spectrum partially regulates the models we consider, often rendering some of the computed quantities finite~\cite{ameliorate}.  It is now clear from the discussion above that to fully regulate this class of models we should in general consider sixth-order ``trans-Bogoliubov''  dispersion relations
\begin{equation}
\omega(\vec k) = \sqrt{ m^2 + c_s^2\; (\vec k)^2 +\Xi^2\; [(\vec k)^2]^2 + [(\vec k)^2]^3 },
\end{equation}
or more schematically
\begin{equation}
\omega(k) = \sqrt{ m^2 + c_s^2\; k^2 +\Xi^2\; k^4 + k^6 }.
\end{equation}
From the point of view of condensed matter and AMO systems such a ``trans-Bogoliubov'' dispersion relation would merely be an artificial regulator; 
however in the context of this present article one might perhaps prefer to view the $k^6$ term as fundamental physics. 

\section{Implications for quantum gravity}
 
While the background physics underlying this article is firmly based in fundamental quantum field theory~\cite{Schweber, Nishijima, Ramond, Rivers, Weinberg},  and ideas from the ``analogue spacetime'' programme~\cite{analogue}, a key stimulus to writing up these observations was the recent article by Petr Ho\v{r}ava~\cite{Horava}, outlining the development of a quantum field theory for 3+1 dimensional gravity --- a theory that is based on a fundamental violation of Lorentz invariance. In that model, Lorentz invariance, and Einstein--Hilbert gravity, is recovered only in the low-momentum (low-spatial-curvature) limit. 

To quickly get to the essence of the argument, I will adopt ``synchronous gauge'' ($N=1$, $N^i=0$), wherein the lapse and shift are trivial and all the physics of the gravitational field is encoded in the spatial metric.  Technically the key step is to consider a model for gravity that is second-order in the time derivatives  of the spatial metric, and that is $(2z)^{th}$-order in the spatial derivatives. The reason this works is that ultimately the spatial Riemann tensor can be written as an infinite-order perturbative expansion around flat 3-space. Schematically, (suppressing all spatial tensor indices) we may write
\begin{equation}
\hbox{Riemann}{(g_{ij}=\delta_{ij}+ h_{ij})}   \sim \sum_{n=0}^\infty \; h^n \; (\nabla^2 h  + \nabla h \cdot \nabla h ).
\end{equation}
But then ``potential'' terms, such as $(\hbox{Riemann})^z$, contain exactly $2z$ spatial derivatives and arbitrary powers of $h$, while the ``kinetic'' term, depending on the square of extrinsic curvature $K$,  is 
\begin{equation}
K^2 \sim \dot h^2.
\end{equation}
Thus an action which is geometrically of the form
\begin{equation}
S \sim \int \left\{ K^2 + (\hbox{Riemann})^z + \dots \right\} \;\d t \,\d x,
\end{equation}
is, from a perturbative point of view, of the form
\begin{equation}
S \sim \int \left\{ \dot h^2 + P(\nabla^{2z},h)\right\} \;\d t \, \d x,
\end{equation}
where $P(\nabla^{2z},h)$  is now an infinite-order polynomial in $h$, which contains up to $2z$ spatial derivatives.  
Viewed as a flat-space quantum field theory, this is thus qualitatively very similar to what I have called $P(\phi)^z_{d+1}$.  

By the dimensional analysis arguments in section \ref{S:power-count},  we see that for $d=z$ the field $h$ is dimensionless, and by power-counting the resulting quantum field theory is then expected to be finite ---  where this means finite in the sense of being both physically well-defined and finite as long as one does not let the Lorentz violation scale go to infinity.  Keeping the Lorentz violation scale finite is now a perfectly sensible thing to do because the regularization has not undermined the internal logical consistency of the quantum field theory. (Of course, for gravity a more careful analysis would need to keep track of all the tensor indices. Furthermore in a general gauge one is dealing with not only the spatial metric, but also the shift vector and lapse function, so that some technical details of the argument will be rather different. Nevertheless, the above argument is the key to understanding why Ho\v{r}ava's model has any hope of being a finite model for quantum gravity.) 

Note that Ho\v{r}ava specifically worked in 3+1 dimensions with a ``potential'' that contained up to six spatial derivatives~\cite{Horava}, as in section~\ref{S:3+1} above. (Ho\v{r}ava's ``potential'' was also constrained by what he called a ``detailed balance'' symmetry~\cite{Horava}.)  From a power counting perspective, as outlined above, it appears likely that Ho\v{r}ava's ideas can be generalized to $d+1$ dimensional gravity, possibly without any need for his ``detailed balance'' condition.

\section{Discussion and conclusions}
 
In summary, in the present article I have described, in I hope a simple and transparent manner,  the use of Lorentz symmetry breaking as an ultraviolet regulator for scalar quantum field theories. Combining power-counting arguments with technical results in Lorentz-violating quantum field theories~\cite{Anselmi}, two key technical results are:
\begin{itemize}
\item 
With normal ordering,  $:\!P(\phi)^{z=d}_{d+1}\!:$ is perturbatively ultraviolet finite.
\item
Even without normal ordering, $P(\phi)^{z>d}_{d+1}$ is perturbatively ultraviolet finite.
\end{itemize}
While Lorentz breaking regulators are computationally difficult to work with, they have the very powerful advantage that they do not damage the physical foundations and internal logical consistency of the underlying theory. This may have applications with regard to developing a tractable quantum field theory whose low-energy limit is Einstein--Hilbert gravity. 

\acknowledgments

I wish to thank Thomas Sotiriou and Silke Weinfurtner for their comments and feedback. 
This research was supported by the Marsden Fund administered by the Royal Society of New Zealand.

\vfill




\begin{thebibliography}{99}
\bibitem{Dirac}
See for instance, the famous (infamous?) comment by Paul Dirac:\\
``I must say that I am very dissatisfied with the situation because this so-called `good theory' does involve neglecting infinities which appear in its equations, neglecting them in an arbitrary way. This is just not sensible mathematics. Sensible mathematics involves neglecting a quantity when it is small --- not neglecting it just because it is infinitely great and you do not want it!''
\end{thebibliography}
\end{document}